\documentclass[twocolumn,amsmath,amssymb,prl,showpacs]{revtex4}

\usepackage{graphicx}
\usepackage{dcolumn}
\usepackage{bm}

\begin{document}
\bibliographystyle{aip}

\title{Sufficient conditions for thermal rectification in graded materials}

\author{Emmanuel Pereira}
 \email{emmanuel@fisica.ufmg.br}
\affiliation{Departamento de F\'{\i}sica--ICEx, UFMG, CP 702,
30.161-970 Belo Horizonte MG, Brazil }

\date{\today}

\begin{abstract}
We address a fundamental problem for the advance of phononics: the
search of a feasible thermal diode. We establish sufficient
conditions for the existence of thermal rectification in general
graded materials. By starting from simple assumptions satisfied by
the usual anharmonic models that describe heat conduction in
solids, we derive an expression for the rectification. The
analytical formula shows how to increase the rectification, and
the conditions to avoid its decay with the system size, a problem
present in the recurrent model of diodes given by the sequential
coupling of two or three different parts. Moreover, for these
graded systems, we show that the regimes of non-decaying rectification and of normal conductivity
 do not overlap. Our
results indicate the graded systems as optimal materials for a
thermal diode, the basic component of several devices of
phononics.

\end{abstract}

\pacs{05.70.Ln; 05.40.-a; 44.10.+i}

\maketitle

\section{Introduction}

The study of the macroscopic laws of thermodynamic transport from
the underlying microscopic models is still a challenge in
statistical physics. In particular, the investigation and control
of the energy transport, which mainly involves conduction of heat
or electricity, is a fundamental problem of huge theoretical and
practical interest. The invention of transistor and other devices used
to control the electric charge flow has led to the well known
development of modern electronics. Its much less
developed counterpart - the study and control of heat current -
has, recently, presented interesting progress, promising to
establish, in addition to electronics, a new  physical branch in
energy and information processing - the phononics \cite{BLiW,
Casati1}: researchers have proposed nanodevices such as thermal
diodes or rectifiers \cite{Casati2, BLi1, BHu1} (already built in
practice \cite{Chang}), thermal transistors \cite{BLi2}, thermal
logic gates \cite{BLi3} and memories \cite{BLi4}. The most
fundamental component of these instruments is the
thermal diode, a device in which heat flows preferably in one
direction. In a short analysis, we may say that this promising advance of
phononics is directly dependent on the  development of its basic component: a
thermal diode with suitable properties.

There are analytical attempts to investigate the phenomenon of
thermal rectification such as the works on spin-boson junctions
\cite{Segal1}, billiard systems \cite{EM}, etc., but most of the
results are by means of computer simulations, see e.g. the work of
B. Li and collaborators \cite{BLi95}. The most common and
recurrent design of diodes is given by the sequential coupling of
two or more chains with different anharmonic potentials
\cite{Casati2, BLi1, BHu1}. Although frequently studied, this
procedure is criticized \cite{BHu1} due to the difficulty to
construct such diode in practice, and due to the significative
decay of the rectification with the system size. Recently, a
different procedure was considered by Chang et al. \cite{Chang},
who built, in an experimental work, the first microscopic
solid-state thermal rectifier  by using a graded material:
 nanotubes externally and inhomogeneously mass-loaded with heavy molecules.
 It is worth to recall that graded materials, i.e.,  inhomogeneous systems whose composition and/or
structure change gradually in space, are abundant in nature, can
also be manufactured,  and have attracted great interest in many
areas \cite{Huang}: there are many works devoted to the study of the electric,
optical, mechanical and other properties of graded materials, but there are few studies in relation
to their heat conduction investigation.

In the present work,  we address this fundamental problem of
phononics: the built of an appropriate thermal diode, namely, a
simple system that may be constructed in practice, and with a
rectification that does not decay with the system size. We start
from simple conditions for the local thermal conductivity,
conditions that are quite general and that are satisfied by
anharmonic crystal models used to describe heat conduction in
solids, and then we show that they are sufficient to lead to
rectification in graded models. Moreover, we derive an expression
for such rectification that allows us to see how to make it
larger, and how to avoid its decay with the system size. In short,
we show that  properly manipulated graded materials have suitable
properties of rectification, and so, they  shall play a central
role in the building of thermal nano-devices. The simplicity of
the initial conditions and of the arguments to establish the
results shows the ubiquity of thermal rectification in graded
systems. Moreover, the existence of simple ingredients for the
rectification, as described here, deserves attention: as well
known, in the literature, the mechanism behind
 rectification in graded models is far from being clear. E.g., we  recall the comment of G.
 Casati \cite{Casati1} on the explanation of Chang. et al \cite{Chang}: ``the authors
 speculate that solitons might be involved in the rectification process, but this is
 still to be confirmed''. Here, we do not have to make any speculation  about the
 vibrational spectra or other intricate property.

\section{Existence of Thermal Rectification}

Let us introduce our assumptions and derive our results.

We consider a chain with $N$ sites, where the first site is
connected to a thermal bath at temperature $T_{1}$, and the last
site is connected to a bath at temperature $T_{N}$. It is possible
to extend our analysis also for a $d$-dimensional lattice with two
thermal baths at the boundaries: the chain structure is
represented by the axis (direction) of the heat flow. We assume
that it is possible to build a temperature gradient in the system.
Such condition always happens if the Fourier's law holds, but we
do not demand  this law here (anyway, we will study cases where
Fourier's law holds). Precisely, we assume that the heat flow from
site $j$ to $j+1$ is given by
\begin{equation}
\mathcal{F}_{j,j+1} = -\mathcal{K}_{j}(\nabla T)_{j} = \frac{1}{\mathcal{C}_{j}T_{j}^{\alpha}+
\mathcal{C}_{j+1}T_{j+1}^{\alpha}} \left( T_{j}-T_{j+1} \right) ,\label{fluxo}
\end{equation}
i.e., with, say, the local thermal conductivity given by the
average of a function of the local temperatures and other
parameters of the system. For the homogeneous model, such
expression reads
\begin{equation*}
\mathcal{F}_{j,j+1}^{H} = \frac{1}{\mathcal{C}(T_{j}^{\alpha}+
T_{j+1}^{\alpha})} \left( T_{j}-T_{j+1} \right)=
\frac{1}{\mathcal{C}'\bar{T}_{j}^{\alpha}} \left( T_{j}-T_{j+1}
\right) ,
\end{equation*}
where $\bar{T}_{j}^{\alpha} = \left(
T_{j}^{\alpha}+T_{j+1}^{\alpha} \right)/2$, $\mathcal{C}'=
2\mathcal{C}$, which is exactly the  formula described by several
results (on homogeneous models) from the literature: e.g., in
ref.\cite{LS}, we have $\alpha = 2$, $\mathcal{C}'T^{2} =
1/\mathcal{K} = \lambda^{2}T^{2}/\omega^{9}\mu^{3}$, where
$\lambda$ is the coefficient of the quartic anharmonic potential,
$\omega$ is the coefficient of the interparticle quadratic
interaction, and $\mu$ is the harmonic pinning coefficient. Still
for this $\phi^{4}$ model, in different conditions and methods,
Bricmont and Kupiainen, \cite{Kup} and Spohn et al. \cite{Spohn}
found $\mathcal{K}_{j}= T_{j}^{-2}$. And, in reference to works
with detailed computer simulations, Aoki and Kusnezov \cite{AK}
obtain for this one-dimensional $\phi^{4}$ model, $\mathcal{K}
\propto T^{-1,35}$; similarly, N. Li and B. Li \cite{LiLi} obtain
$\mathcal{K} \propto T^{-1,5}$, with slight changes in  the
exponent that depend on the values of the pinning and
anharmonicity.
It is also worth to recall that, by using an analytical simplified
scheme (derived from a rigorous and more intricate approach
\cite{PF}), we obtain a similar formula for the local thermal
conductivity of the graded anharmonic self-consistent chain
\cite{Prapid}, i.e., of the anharmonic, inhomogeneous model given
by a chain of oscillators with quartic on-site potential,
quadratic nearest-neighbor interparticle interaction, particles
with different masses and inner stochastic reservoirs connected to
each site.

Let us now prove the existence of thermal rectification for a
graded anharmonic system with a temperature gradient in the bulk,
 and whose local
thermal conductivity depends on temperature (which does not follow
in the harmonic case), and changes as we run the chain.

From the fact that the heat current comes into the system by the first site,
passes trough the chain and goes out by the last site, we have
\begin{equation}
\mathcal{F}_{1,2} = \mathcal{F}_{2,3} = \ldots = \mathcal{F}_{N-1,N} \equiv \mathcal{F} \label{corrente}.
\end{equation}
These equations together with eq.(\ref{fluxo}) give us
\begin{eqnarray*}
\mathcal{F}(\mathcal{C}_{1}T_{1}^{\alpha} + \mathcal{C}_{2}T_{2}^{\alpha}) &=& T_{1} - T_{2} \\
\mathcal{F}(\mathcal{C}_{2}T_{2}^{\alpha} + \mathcal{C}_{3}T_{3}^{\alpha}) &=& T_{2} - T_{3} \\
\ldots &=& \ldots \\
\mathcal{F}(\mathcal{C}_{N-1}T_{N-1}^{\alpha} + \mathcal{C}_{N}T_{N}^{\alpha}) &=& T_{N-1} - T_{N} .
\end{eqnarray*}
Summing up the equations, we find
\begin{equation*}
\mathcal{F} = \mathcal{K}\frac{(T_{1}-T_{N})}{N-1} ,
\end{equation*}
where
\begin{eqnarray}
\lefteqn{\mathcal{K} = \left\{ \mathcal{C}_{1}T_{1}^{\alpha} +
2\mathcal{C}_{2}T_{2}^{\alpha} + \right.\ldots} \nonumber \\
&& \left. + 2\mathcal{C}_{N-1}T_{N-1}^{\alpha} +
\mathcal{C}_{N}T_{N}^{\alpha}\right\}^{-1}\cdot (N-1)
,\label{condutividade}
\end{eqnarray}
that is the Fourier's law for the case of the thermal conductivity
$\mathcal{K}$ remaining finite as $N\rightarrow\infty$. From
eq.(\ref{fluxo}) and eq.(\ref{corrente}), it follows that
\begin{eqnarray}
\frac{T_{1}-T_{2}}{\mathcal{C}_{1}T_{1}^{\alpha}+\mathcal{C}_{2}T_{2}^{\alpha}}
&=&
\frac{T_{2}-T_{3}}{\mathcal{C}_{2}T_{2}^{\alpha}+\mathcal{C}_{3}T_{3}^{\alpha}}
= \ldots \nonumber \\
&=& \frac{T_{N-1}-T_{N}}{\mathcal{C}_{N-1}T_{N-1}^{\alpha}+
\mathcal{C}_{N}T_{N}^{\alpha}} \label{tibau}.
\end{eqnarray}
Thus, given the temperatures of the baths $T_{1}$ and $T_{N}$, by using the equations above
we determine the inner temperatures $T_{2}$, $T_{3}$, \ldots, $T_{N-1}$.
For ease of computation, let us consider the system submitted to a small gradient of temperature: $T_{1} = T + a_{1}\epsilon$,
$T_{N} = T + a_{N}\epsilon$, $\epsilon$ small. Hence, $T_{k} = T + a_{k}\epsilon + \mathcal{O}(\epsilon^{2})$. We will carry out the computations only up to $\mathcal{O}(\epsilon)$.
And so, up to $\mathcal{O}(\epsilon)$, we have $T_{k}^{\alpha} = T^{\alpha} + \alpha T^{\alpha -1}\epsilon a_{k}$ (that comes from the Taylor series), and
$$
\frac{T_{k}-T_{k+1}}{\mathcal{C}_{k}T_{k}^{\alpha}+\mathcal{C}_{k+1}T_{k+1}^{\alpha}} =
\frac{\left(a_{k}-a_{k+1}\right)\epsilon}{\left( \mathcal{C}_{k} + \mathcal{C}_{k+1}\right)T^{\alpha}} ,
$$
as said, up to $\mathcal{O}(\epsilon)$.
From this equation and eq.(\ref{tibau}), we obtain
\begin{equation}
\frac{a_{1}-a_{2}}{\mathcal{C}_{1}+\mathcal{C}_{2}} = \frac{a_{2}-a_{3}}{\mathcal{C}_{2}+\mathcal{C}_{3}} = \ldots
= \frac{a_{N-1}-a_{N}}{\mathcal{C}_{N-1}+ \mathcal{C}_{N}} \label{eqa}.
\end{equation}
We may rewrite these equations as
\begin{eqnarray*}
\frac{a_{1}-a_{2}}{\mathcal{C}_{1}+\mathcal{C}_{2}} &=& \frac{a_{1}-a_{2}}{\mathcal{C}_{1}+\mathcal{C}_{2}} \\
\frac{a_{1}-a_{2}}{\mathcal{C}_{1}+\mathcal{C}_{2}} &=& \frac{a_{2}-a_{3}}{\mathcal{C}_{2}+\mathcal{C}_{3}} \\
\ldots &=& \ldots \\
\frac{a_{1}-a_{2}}{\mathcal{C}_{1}+\mathcal{C}_{2}} &=& \frac{a_{N-1}-a_{N}}{\mathcal{C}_{N-1}+\mathcal{C}_{N}} .
\end{eqnarray*}
Summing them up, we obtain
\begin{eqnarray*}
\frac{a_{1}-a_{2}}{\mathcal{C}_{1}+\mathcal{C}_{2}} & \cdot & \left(\mathcal{C}_{1} + 2\mathcal{C}_{2} + \ldots + 2\mathcal{C}_{N-1} + \mathcal{C}_{N}\right) = a_{1}-a_{N} \\
&&   \Rightarrow  a_{2} = a_{1} + \frac{\left( a_{N}-a_{1} \right)}{\tilde{\mathcal{C}}(N)}\left(\mathcal{C}_{1}+\mathcal{C}_{2}\right) ,
\end{eqnarray*}
where $\tilde{\mathcal{C}}(N) \equiv (\mathcal{C}_{1} +
2\mathcal{C}_{2} + \ldots + 2\mathcal{C}_{N-1} +
\mathcal{C}_{N})$. Similarly, writing
$(a_{k-1}-a_{k})/(\mathcal{C}_{k-1}+\mathcal{C}_{k})$ instead of
$(a_{1}-a_{2})/(\mathcal{C}_{1}+\mathcal{C}_{2})$ in the LHS of the
list of equations above, we obtain
\begin{equation}
a_{k} = a_{1} + \frac{\left( a_{N}-a_{1} \right)}{\tilde{\mathcal{C}}(N)} \tilde{\mathcal{C}}(k) ,
\end{equation}
for $k=2, \ldots, N-1$.
And so, for the thermal conductivity (\ref{condutividade}) it
follows that
\begin{eqnarray}
 \mathcal{K} &=& (N-1) \cdot \left\{ T^{\alpha}\tilde{\mathcal{C}}(N)
+ \alpha T^{\alpha -1}\epsilon\left( a_{1}\mathcal{C}_{1} +
2a_{2}\mathcal{C}_{2} + \ldots \right.\right. \nonumber \\
&& \left.\left. + 2a_{N-1}\mathcal{C}_{N-1} + a_{N}\mathcal{C}_{N}
\right) \right\}^{-1} .
\end{eqnarray}

To investigate the existence or absence of rectification, we need
to analyze the heat flow for the system with inverted thermal baths,
that is, we compute the new thermal conductivity for the same
system, but with temperatures $T'$, where $T'_{1} = T_{N}$ and
$T'_{N} = T_{1}$. Following the previous manipulations, we see
that, in the system with inverted baths, the new temperature for
the site $k$ is $T'_{k} = T + a'_{k}\epsilon$, where, for $k = 2,
3, \ldots, N-1$,
\begin{equation}
a'_{k} = a_{N} - \frac{\left( a_{N}-a_{1}
\right)}{\tilde{\mathcal{C}}(N)} \tilde{\mathcal{C}}(k) .
\end{equation}
Obviously: $a'_{1}=a_{N}$, and $a'_{N}=a_{1}$. Hence, the
expression for the ``inverted'' thermal conductivity becomes
\begin{eqnarray}
 \mathcal{K}' &=& (N-1) \cdot \left\{ T^{\alpha}\tilde{\mathcal{C}}(N)
+ \alpha T^{\alpha -1}\epsilon\left( a'_{1}\mathcal{C}_{1} +
2a'_{2}\mathcal{C}_{2} + \ldots \right.\right. \nonumber \\
&& \left.\left. + 2a'_{N-1}\mathcal{C}_{N-1} +
a'_{N}\mathcal{C}_{N} \right) \right\}^{-1} .
\end{eqnarray}
And, with simple manipulations, we get
\begin{eqnarray}
\frac{1}{\mathcal{K}} - \frac{1}{\mathcal{K}'} &=&
\frac{\alpha T^{\alpha -1}\epsilon (a_{1}-a_{N})}{(N-1)\tilde{\mathcal{C}}(N)} \nonumber \\
 && \times \left\{ \tilde{\mathcal{C}}(N)^{2} - 4\tilde{\mathcal{Q}}(N) -
2\mathcal{C}_{N}\tilde{\mathcal{C}}(N) \right\} ,
\end{eqnarray}
where $\tilde{\mathcal{Q}}(N) \equiv \tilde{\mathcal{C}}(2)\mathcal{C}_{2} + \ldots +
\tilde{\mathcal{C}}(N-1)\mathcal{C}_{N-1}$. As a simple test for the expression above, note that it vanishes (as expected) in
the case of a homogeneous system ($\mathcal{C}_{1}= \mathcal{C}_{2} = \ldots = \mathcal{C}_{N}$).

To continue the analysis, we take a chain with three sites (say, the smallest possible system). A direct computation gives us
\begin{equation*}
(\tilde{\mathcal{C}}(3))^{2} - 4\tilde{\mathcal{Q}}(3) - 2\mathcal{C}_{3}\tilde{\mathcal{C}}(3) = \mathcal{C}_{1}^{2} - \mathcal{C}_{3}^{2} .
\end{equation*}
Now we prove, by induction, that such relation is valid for any number of sites: we assume that it is valid for $k$ sites (i.e., for $k$ replacing 3 in the relation above),
and then we show that it follows for $k+1$. In fact, by using the definitions
we see that
\begin{eqnarray*}
\tilde{\mathcal{C}}(k+1) &=& \mathcal{C}_{1} + 2\mathcal{C}_{2} +
\ldots + 2\mathcal{C}_{k} + \mathcal{C}_{k+1} \\
 &=& \tilde{\mathcal{C}}(k) + \mathcal{C}_{k} +
\mathcal{C}_{k+1} ,\\
\tilde{\mathcal{Q}}(k+1) &=& \tilde{\mathcal{C}}(2)\mathcal{C}_{2} + \ldots + \tilde{\mathcal{C}}(k)\mathcal{C}_{k} =
\tilde{\mathcal{Q}}(k)  + \tilde{\mathcal{C}}(k)\mathcal{C}_{k} .
\end{eqnarray*}
Then, a direct computation shows that
\begin{equation*}
(\tilde{\mathcal{C}}(K+1))^{2} - 4\tilde{\mathcal{Q}}(k+1) - 2\mathcal{C}_{k+1}\tilde{\mathcal{C}}(k+1) = \mathcal{C}_{1}^{2} - \mathcal{C}_{k+1}^{2} .
\end{equation*}
Hence, for the difference between the thermal conductivities of the system with $N$ sites, we obtain
\begin{equation}
\frac{1}{\mathcal{K}} - \frac{1}{\mathcal{K}'} = \frac{\alpha
T^{\alpha -1}\epsilon (a_{1}-a_{N})}{(N-1)\tilde{\mathcal{C}}(N)}
\left[ \mathcal{C}_{1}^{2} - \mathcal{C}_{N}^{2} \right] ,
\end{equation}
where, we recall, $\alpha T^{\alpha -1}\epsilon (a_{1}-a_{N})$ in
the numerator above is $T_{1}^{\alpha}-T_{N}^{\alpha}$ up to
$\mathcal{O}(\epsilon)$. Thus, the existence of thermal
rectification for anisotropic, e.g. graded,  materials is
transparent.

\section{Rectification Properties}

Now, let us examine the rectification in details and search for
conditions leading to suitable properties. First, we write the
expression for the rectification factor $f_{r}$
\begin{equation*}
f_{r}  \equiv  \frac{|\mathcal{K}-\mathcal{K}'|}{\mathcal{K}'}
      \approx  \frac{|T_{1}^{\alpha}-T_{N}^{\alpha}|}{T^{\alpha}}
     \frac{|\mathcal{C}_{1}^{2}-\mathcal{C}_{N}^{2}|}{(\tilde{\mathcal{C}}(N))^{2}}
     .
\end{equation*}
Hence, fixed the temperatures at the boundaries, the behavior of
$f_{r}$ with $N$ is given by
$|\mathcal{C}_{1}^{2}-\mathcal{C}_{N}^{2}|/(\tilde{\mathcal{C}}(N))^{2}$.
We recall that
$$
\tilde{\mathcal{C}}(N) = \mathcal{C}_{1} + 2\mathcal{C}_{2} +
\ldots +  2\mathcal{C}_{N-1} + \mathcal{C}_{N} \approx
2\int_{1}^{N}\mathcal{C}_{x} dx .
$$
 And, for a small
gradient of temperature in the system,
$$
\mathcal{K} = (N-1)/\{T^{\alpha}\tilde{\mathcal{C}}(N) +
\mathcal{O}(\epsilon)\}.
$$
 Thus, to get a normal conductivity
(Fourier's law) we must have $\tilde{\mathcal{C}}(N) \sim N$,
i.e., $\mathcal{C}_{N} \sim$  constant. That is, for these graded
systems, at least at small temperature gradients, if the
conductivity is normal then the rectification factor decays to
zero as $N\rightarrow\infty$. To avoid the decay of the
rectification factor, for example, to make it finite and nonzero
as $N\rightarrow\infty$, we need to take $\mathcal{C}_{N} \sim
c\exp(\gamma N)$. And so, $\tilde{\mathcal{C}}(N) \sim
c(\exp(\gamma N)-1)/\gamma$. For $\gamma>0$,
$\tilde{\mathcal{C}}(N)$ has exponential growth and
$\mathcal{K}(N)\rightarrow 0$ as $N\rightarrow\infty$. For
$\gamma<0$, $\tilde{\mathcal{C}}(N)\rightarrow$ constant and
$\mathcal{K} \sim N$, i.e., we have an abnormal conductivity. That
is, the regimes of non-decaying rectification and of normal
conductivity do not overlap. The possibility of a non-decaying
rectification is a very important property: as recalled before,
the decay of rectification is a problem for the usual diodes given
by the sequential coupling of different parts.

Moreover, still from the previous expression (take $T_{1}>T_{N}$
and $\mathcal{C}_{N}> \mathcal{C}_{1}$), we see that the thermal
conductivity is smaller when the heat flows from the sites with
larger $\mathcal{C}$ to the sites with smaller $\mathcal{C}$.

In short, we have shown that in a lattice system where it is
possible to build a temperature gradient, i.e., with the heat flow
from site $j$ to $j+1$ given by eq.(\ref{fluxo}), with graded
structure (i.e. graded $\mathcal{C}_{j}$) and with local thermal
conductivity dependent on temperature (see eq.(\ref{fluxo})), we
will always have thermal rectification. To be precise, we need to
recall that in our proof, for ease of computation, we have assumed
a system with small temperature gradient (however, we believe that
it is not a necessary condition - more comments ahead). It is
interesting to note that such conditions - temperature gradient in
the bulk, local conductivity dependent on temperature and a graded
structure - appear in the quantum harmonic self-consistent chain
of oscillators \cite{PPLA}, a system that presents rectification
in opposition to its classical version (with a conductivity that
does not depend on temperature).

To give a concrete example, we turn to the chain with homogeneous anharmonic
potential, homogeneous interparticle interactions, etc, but with graded masses. For the model with
inner self-consistent reservoirs, weak nearest neighbor interactions, quartic anharmonicity, in
an approximate calculation \cite{Prapid}, we have
\begin{equation*}
\mathcal{F}_{j,j+1} = \frac{C}{(m_{j+1}T_{j}^{1/2}+
m_{j}T_{j+1}^{1/2})} \left( T_{j}-T_{j+1} \right) ,
\end{equation*}
where $C$ involves the coefficients for the anharmonicity, interparticle interaction, etc. The
denominator of the expression above may be written as $[(T_{j}^{1/2}/\rho_{j,j+1}) + (T_{j+1}^{1/2}/\rho_{j+1,j})]$,
where $\rho_{j,j+1} = m_{j}/m_{j+1}m_{j}$,  $\rho_{j+1,j} = m_{j+1}/m_{j+1}m_{j}$. To follow, we define
$$
\bar{\rho}_{j} \equiv \frac{\rho_{j,j-1} + \rho_{j,j+1}}{2} = \frac{1}{2}\cdot \frac{(m_{j+1} + m_{j-1})}{(m_{j-1}m_{j+1})} ,
$$
 i.e., $\bar{\rho}_{j}$ is
proportional to the inverse of a reduced mass. Hence, considering
the entire system $j=1, \ldots, N$,  we approximately have
$\mathcal{F}_{j,j+1}$ given by eq.(\ref{fluxo}) with
$\mathcal{C}_{j} = 1/\bar{\rho}_{j}C$. And the analysis follows as
previously described: now with the bigger flow in the direction
from the larger to the smaller masses. It is worth to recall that
such property, a bigger heat flow from the larger to smaller
densities, as described here, has been already experimentally
described \cite{Chang}.

Similar properties appear in a system with homogeneous
particle masses, but graded anharmonic on-site potentials or
graded interpaticle interactions.

\section{Final Remarks}

We have some remarks. First, we stress that we have presented here
sufficient, not necessary, conditions for manifesting thermal
rectification in anisotropic systems. In ref.\cite{LiR}, by
computer simulations, the authors describe rectification in a
graded mass Fermi-Pasta-Ulam chain, a model with an invariant
translational potential and  abnormal conductivity (even for the
case of homogenous masses). We also recall that, for the (very
different) case of a system of two-terminal junctions, sufficient
conditions for rectification have been described in a recent work
by Wu and Segal \cite{Segal2}.

A further investigation of great interest is the behavior of the
graded system as submitted to a large gradient of temperature: we
believe that it shall lead to a significative rectification. In
ref.\cite{CMP}, for some specific models given by chaotic billiard
systems, the authors claim that there is a significative
rectification ``provided the temperatures (of the two sides of the
system) are strongly different..."

To conclude, we emphasize that due to their simplicity, the
assumptions and arguments described here follow for many of the
usual systems modeling heat conduction in solids: it shows the
ubiquity of rectification in graded systems. Moreover,  the
existence of simple conditions for the existence of an efficient
rectification, and the fact that graded systems may be constructed
in practice (and are even abundant in nature) indicate that they
are optimal material to be used in the construction of a thermal
diode (and also thermal transistors, etc), and so, their use shall
certainly contribute to the advance of phononics.

Work supported by CNPq (Brazil).

\end{document}